# Semi-analytical solution for the trapped orbits of satellite near the planet in ER3BP

**Sergey Ershkov***, Affiliation[1]: Plekhanov Russian University of Economics,

Scopus number 60030998, e-mail: sergej-ershkov@yandex.ru

Affiliation[2]: Sternberg Astronomical Institute, M.V. Lomonosov's Moscow State University, 13 Universitetskij prospect, Moscow 119992, Russia,

**Alla Rachinskaya**, Odessa I. I. Mechnikov National University,

2 Dvoryanskaya St., Odessa, Ukraine, e-mail: rachinskaya@onu.edu.ua

## Abstract

In this paper, we present a new ansatz for solving equations of motion for the *trapped orbits* of the infinitesimal mass (satellite), which is locked in the space trap to be moving near the planet in case of the elliptic restricted problem of three bodies, ER3BP (with Keplerian *elliptic* trajectories of primaries Sun and planet around each other). A new type of the solving procedure is implemented here to obtain the coordinates of the infinitesimal mass (satellite) with its orbit located near the planet. The system of equations of motion was applied for obtaining of the semi-analytic and analytic solutions.

It is obtained that two cartesian coordinates (in a plane of mutual rotation of primaries Sun and planet around each other) depend on the true anomaly and a function which determines the quasi periodic character of solution, while the third coordinate (perpendicular to the plane of rotation of primaries) is quasi-periodically varying with true anomaly.

## 1. Introduction.

In the restricted three-body problem (R3BP), the equations of motion describe the dynamics of an infinitesimal mass $m$ under the action of gravitational forces effected by two celestial bodies of giant masses $M_{Sun}$ and $m_{planet}$ ($m_{planet} < M_{Sun}$), which are rotating around their common centre of mass on Keplerian trajectories. The small mass $m$ (satellite) is supposed to be moving as first approximation inside of *restricted* region of space near the planet of mass $m_{planet}$ or inside of so-called *Hill sphere* [1] radius:

$$r_H \cong a_p \cdot (1 - e^2) \cdot \left( \frac{m_{planet}}{3(M_{Sun} + m_{planet})} \right)^{\frac{1}{3}}$$

where $a_p$ is semimajor axis of the planet's orbit, $e$ is the eccentricity of it's orbit.

It is worth noting that there is a large number of previous and recent works concerning analytical development with respect to the R3BP equations which should be mentioned accordingly [2-5].

We should especially emphasize the theory of orbits, which was developed in profound work [5] by V. Szebehely for the case of the circular restricted problem of three bodies (CR3BP) (primaries $M_{Sun}$ and $m_{planet}$ are rotating around their common centre of mass on *circular* orbits) as well as the case of the elliptic restricted problem of three bodies (ER3BP), where the primaries $M_{Sun}$ and $m_{planet}$ are rotating around their common centre of mass on *elliptic* orbits.

Unlike the CR3BP [6], the position of the primaries is not fixed in the rotating frame as they move along elliptical orbits: their relative distance $\rho$ is not constant in time

$$\rho = \frac{a_p}{1 + e \cdot \cos f}$$

where $f$ is the true anomaly (the unit of distances is chosen so that $a_p = 1$).

As for the purpose of the current research, we can formulate it as follows: the main aim is to find a kind of the semi-analytical solution to the system of equations under consideration. Namely, each exact or even semi-analytical solution can clarify the structure, intrinsic code and topology of the variety of possible solutions (from mathematical point of view); here, exact or semi-analytical solution solution should be treated not only as analytical formulae in quadratures, but a system of ordinary differential equations (each for one appropraite variable) with well-known code for analytical or numerical resolving to be presented in their final form.

## 2. Mathematical model, equations of motion.

According to [6-7], in the ER3BP equations of motion of the infinitesimal mass *m* (satellite) can be represented in the synodic co-rotating frame of a Cartesian coordinate system $\vec{r} = \{x, y, z\}$ in non-dimensional form (*at given initial conditions*):

$$\ddot{x} - 2\dot{y} = \frac{\partial\Omega}{\partial x},$$

$$\ddot{y} + 2\dot{x} = \frac{\partial\Omega}{\partial y}, \qquad (1)$$

$$\ddot{z} = \frac{\partial\Omega}{\partial z},$$

$$\Omega = \frac{1}{1+e\cdot\cos f}\left[\frac{1}{2}\left(x^2+y^2-z^2\cdot e\cdot\cos f\right)+\frac{(1-\mu)}{r_1}+\frac{\mu}{r_2}\right], \qquad (2)$$

where dot indicates (d/d*f*) in (1), Ω is the scalar function, and

$$r_1^2 = (x-\mu)^2 + y^2 + z^2,$$

$$r_2^2 = (x-\mu+1)^2 + y^2 + z^2, \qquad (3)$$

where $r_i$ ($i$ =1, 2) are distances of the infinitesimal mass $m$ from the primaries $M_{Sun}$ and $m_{planet}$, respectively [7].

Now, the unit of mass is chosen in (1) so that the sum of the primary masses is equal to 1. We suppose that $M_{Sun} \cong 1 - \mu$ and $m_{planet} = \mu$, where $\mu$ is the ratio of the mass of the smaller primary to the total mass of the primaries and $0 < \mu \leq 1/2$. The unit of time is chosen so that the gravitational constant is equal to 1 in (2).

We neglect the effect of variable masses of the primaries [8] as well as the effect of their *oblateness* as was considered earlier in [9]. As for the domain where the aforesaid infinitesimal mass $m$ is supposed to be moving, let us consider the Cauchy problem in the whole space. Besides, we should note that the second terms in the left parts of Eqns. (1) are associated with the components of the *Coriolis* acceleration. Finally, let us additionally note that the spatial ER3BP when $e > 0$ and $\mu > 0$ is not conservative, and no integrals of motion are known [7].

## 3. Reduction of the system of equations (1).

Aiming the aforementioned way of constructing the semi-analytical solution, let us present Eqns. (1) in a suitable for analysis form as below by appropriately transforming the right parts with regard to partial derivatives with respect to the proper coordinates $\{x, y, z\}$ [10]

$$\ddot{x} - 2\dot{y} = \frac{1}{1+e\cdot\cos f}\cdot\left[x - \frac{(1-\mu)(x-\mu)}{\left((x-\mu)^2+y^2+z^2\right)^{\frac{3}{2}}} - \frac{\mu(x-\mu+1)}{\left((x-\mu+1)^2+y^2+z^2\right)^{\frac{3}{2}}}\right],$$

$$\ddot{y} + 2\dot{x} = \frac{1}{1+e\cdot\cos f}\cdot\left[y - \frac{(1-\mu)y}{\left((x-\mu)^2+y^2+z^2\right)^{\frac{3}{2}}} - \frac{\mu y}{\left((x-\mu+1)^2+y^2+z^2\right)^{\frac{3}{2}}}\right], \qquad (4)$$

$$\ddot{z} = \frac{1}{1+e\cdot\cos f}\cdot\left[-z\cdot e\cdot\cos f - \frac{(1-\mu)z}{\left((x-\mu)^2+y^2+z^2\right)^{\frac{3}{2}}} - \frac{\mu z}{\left((x-\mu+1)^2+y^2+z^2\right)^{\frac{3}{2}}}\right],$$

Let us transform system of equations (4) to the form which would be convenient for further analysis. From first and second of Eqns. (4) we obtain ($y \neq 0$):

$$(\ddot{x} - 2\dot{y})\cdot(1+e\cdot\cos f) - x = (x-\mu)\cdot\left\{\frac{(\ddot{y} + 2\dot{x})\cdot(1+e\cdot\cos f) - y}{y}\right\} - \frac{\mu}{\left((x-\mu+1)^2 + y^2 + z^2\right)^{\frac{3}{2}}} \qquad (5)$$

In (5), $z^2$ is given as the function of {$x$, $y$}, their derivatives { $\dot{x}$, $\dot{y}$ } with respect to the $f$, accelerations { $\ddot{x}$, $\ddot{y}$ }, and true anomaly $f$.

But from the second and third of Eqns. (4) it follows ({$y$, $z$} $\neq 0$, { $\dot{x}$, $\dot{y}$, $\dot{z}$ } $\neq 0$):

$$(1+e\cdot\cos f)\left(\frac{\ddot{z}}{z}+1\right) = \frac{(\ddot{y} + 2\dot{x})\cdot(1+e\cdot\cos f)}{y}, \quad \Rightarrow \quad \frac{\ddot{z}}{z} + 1 = \frac{(\ddot{y} + 2\dot{x})}{y} \qquad (6)$$

## 4. Approximated solutions of Eqns. (1)-(3) for the class of *trapped motions*.

Let us assume that coordinates $\vec{r}$ ={$x$, $y$, $z$} of solutions of system (1) belong to the class of *trapped motions* of the infinitesimal mass $m$, which is moving near the planet $m_{planet}$ (taking into account the equality $\rho = \frac{a_p}{1+e\cdot\cos f}$):

$$\frac{|\vec{r}_2|}{|\vec{r}_1|} << 1, \quad |\vec{r}_1| \cong \frac{a_p}{1+e\cdot\cos f} + \delta, \quad |\delta| << a_p \qquad (7)$$

but the aforementioned infinitesimal mass $m$ are, nevertheless, located on each step of its trajectory at a large distance from the Sun ($M_{Sun}$) insofar; here $|\vec{r}_2| > R_p$, whereas $R_p$ is the radius of planet $m_{planet}$.

Thus if we take into consideration the additional restriction (7) with respect to the components of solution in Eqns. (1)-(3), the aforesaid assumption should simplify the third of equations (4) accordingly (except the obvious case $\{z,\ \ddot{z}\} = 0$ in our further analysis):

$$\ddot{z}\cdot(1+e\cdot\cos f)\ +\ z\cdot e\cdot\cos f\ =\ -\ \frac{z}{\left|\vec{r}_2\right|^3}\cdot\left\{(1-\mu)\cdot\frac{\left|\vec{r}_2\right|^3}{\left|\vec{r}_1\right|^3}\ +\ \mu\right\},\quad\Rightarrow$$

$$\left\{\left(\frac{\left|\vec{r}_2\right|}{\left|\vec{r}_1\right|}\right)^3 \to 0\right\}\quad\Rightarrow\quad \left|\vec{r}_2\right| \cong \frac{\mu^{\frac{1}{3}}}{\left(-\ \dfrac{\ddot{z}}{z}\cdot(1+e\cdot\cos f)\ -\ e\cdot\cos f\right)^{\frac{1}{3}}},\quad\Rightarrow$$

$$y\ \cong\ \pm\sqrt{\frac{\mu^{\frac{2}{3}}}{\left(\dfrac{\ddot{z}}{z}\cdot(1+e\cdot\cos f)\ +\ e\cdot\cos f\right)^{\frac{2}{3}}}\ -\ \left(\ (x-\mu+1)^2\ +\ z^2\ \right)},\tag{8}$$

where appropriate restriction should be valid

$$\frac{\mu^{\frac{2}{3}}}{\left(\dfrac{\ddot{z}}{z}\cdot(1+e\cdot\cos f)\ +\ e\cdot\cos f\right)^{\frac{2}{3}}}\ -\ z^2\ \geq\ (x-\mu+1)^2\ ,$$

Thus, we have expressed in (8) the coordinate $y$ *via* coordinates $\{x,\ z\}$ and second derivative of coordinate $\ddot{z}$ with respect to the true anomaly $f$ (as a first approximation).

Now let us present Eqn. (6) in a form of the *Riccati*-type ordinary differential equation [11] for coordinate $z$, depending on the coordinate $y$ and on the appropriate derivatives of coordinates $\{\ \dot{x},\ \ddot{y}\ \}$ with respect to the true anomaly $f$

$$\ddot{z} + \left(1 - \frac{(\ddot{y} + 2\dot{x})}{y}\right)\cdot z = 0 \qquad (9)$$

So, equation (9) should determine the proper *quasi-periodic* solution for coordinate $z$ if the solutions for coordinates $\{x, y\}$ are already obtained.

Let us present further the solutions of Eqn. (9)

$$\left(1 - \frac{(\ddot{y} + 2\dot{x})}{y}\right) = \alpha(f) \;\;\Rightarrow\;\; \ddot{z} + \alpha\cdot z = 0 \qquad (10)$$

The aforementioned presentation of solutions in a form (10) for coordinate $z$ is obviously useful from practical point of view in celestial mechanics for the reason that such the solutions e.g. could be presenting the *quasi-periodical* dependence of coordinate $z$ with respect to the true anomaly $f$ (if α is considered to be slowly varying parameter or *circa constant*, as we can see in our analysis).

The second advance of exploring the differential invariant (6) in a form (10) is that we can reduce one of two equations (second or third) of system (4), which was used at derivation of differential invariant (6). Let us choose the third equation for this aim

$$\alpha = \frac{1}{1 + e\cdot\cos f}\cdot\left[e\cdot\cos f + \frac{(1-\mu)}{\left((x-\mu)^2 + y^2 + (z)^2\right)^{\frac{3}{2}}} + \frac{\mu}{\left((x-\mu+1)^2 + y^2 + (z)^2\right)^{\frac{3}{2}}}\right] \qquad (11)$$

Meanwhile, equality (11) reveals the *quasi-periodic* type of the solutions for Eqns. (10) (if $\alpha$ is considered to be slowly varying parameter or *circa constant*): indeed, taking into account the additional restriction (7), we can make a reasonable conclusion from (11) that α > 0 in any case (for the motions which can be expected according to the additional assumption (7)).

So, in this case Eqn. (10) yields the classical periodic type of solutions for coordinate *z* as presented below (e.g., if $\alpha$ is considered to be *circa constant*)

$$z = C_1 \cos(f \cdot \sqrt{\alpha}) + C_2 \sin(f \cdot \sqrt{\alpha}) \qquad (12)$$

where $\{C_1, C_2\}$ = *const*.

We should note also that Eqns. (10) reveals the obvious *quasi-periodic* character of dependence of coordinate *y* on the derivative of coordinate *x* (with respect to the true anomaly *f*). Indeed, we obtain from (10) for coordinate *y*:

$$\ddot{y} + (\alpha - 1) \cdot y = -2\,\dot{x} \qquad (13)$$

which is, in fact, the *equation of forced oscillations* ([10], example 2.36).

For the sake of simplicity, let us consider in our further analysis the *partial* case $\alpha$ = 1 in formulae (10)-(13). Then, integrating both the parts of Eqn. (13) with respect to the true anomaly *f*, we could use the result of integrating in the transformation of the left part of first equation of system (4).

Thus, we should obtain in result the *non-linear* ordinary differential equation of the second order in regard to the coordinate $x(f)$ in case of the given function (12) for the coordinate *z*, $\alpha$ = 1 (whereas the true anomaly *f* is to be slowly varying independent coordinate). Obviously, such the *non-linear* ordinary differential equation of second order (**Appendix, A1**) could be solved by means of numerical methods only. Similar simple case was investigated first in [1] (the well known Clohessy-Wiltshire equations for relative motion when $e \neq 0$) but without obtaining expression for *y* presented by (8).

Finally, let us note that we should restrict choosing of the obtained solutions for the aforesaid *non-linear* ODE of second order with regard to the coordinate $x(f)$ by taking into account the additional condition for the function $r_2 \to R_p$ (while the magnitude of this function should be exceeding the minimal distances within Roche-lobe's region [4-5] for the planet with mass $m_{planet} = \mu$ around which infinitesimal mass is currently rotating in its trapped motion).

As for expression for the function $\delta(f)$, we can obtained it from (7) as below

$$(x-\mu)^2 + y^2 + z^2 \cong \left(\frac{a_p}{1+e\cdot\cos f}+\delta\right)^2, \quad |\delta| << a_p \quad \Rightarrow$$

$$\delta \cong \sqrt{(x-\mu)^2 + y^2 + z^2} - \left(\frac{a_p}{1+e\cdot\cos f}\right) \qquad (14)$$

where expression for $y$ is given in (8) (where expression for $\frac{\ddot{z}}{z}$ could be expressed from (10)), but expression for $z$ is given in (12), $\alpha = 1$.

## 5. Final presentation of the solution.

Let us present the solution $\vec{r} = \{x, y, z\}$ for the *trapped motion* (7) of the infinitesimal mass $m$ (satellite), which is moving near the planet $m_{planet}$ in the ER3BP (1)-(4):

- The key *non-linear* ordinary differential equation of the second order in regard to the coordinate $x(f)$ in case of the given function (12) for the coordinate $z$ ($\alpha = 1$) is obtained below:

Let us we substitute expression (8) for coordinate $y$, expression (12) for coordinate $z$, and the integrated expression for $\dot{y}$ directly $\rightarrow$ into the first equation of system (4) to obtain the *non-linear* ordinary differential equation of the second order with regard to the unknown coordinate $x(f)$ ($\alpha = 1$, $x_0 = const$):

$$\ddot{x} + 4(x - x_0) = \frac{1}{1 + e \cdot \cos f} \cdot \left[ x - \frac{(1-\mu)(x-\mu)}{\left((x-\mu)^2 + y^2 + z^2\right)^{\frac{3}{2}}} - \frac{\mu(x-\mu+1)}{\left(|\vec{r}_2|\right)^3} \right], \qquad \left\{ |\vec{r}_2| \cong \mu^{\frac{1}{3}} \right\},$$

$$\Rightarrow \qquad \ddot{x} + 4(x - x_0) = -\frac{(1-\mu)}{1 + e \cdot \cos f} \cdot \left[ 1 + \frac{(x-\mu)}{\left((x-\mu)^2 + y^2 + z^2\right)^{\frac{3}{2}}} \right],$$

$$\left\{ y \cong \pm\sqrt{\mu^{\frac{2}{3}} - \left( (x-\mu+1)^2 + z^2 \right)}, \quad \mu^{\frac{2}{3}} - \left( (x-\mu+1)^2 + z^2 \right) > 0 \right\} \qquad \Rightarrow$$

$$\Rightarrow \qquad \ddot{x} + 4(x - x_0) = -\frac{(1-\mu)}{1 + e \cdot \cos f} \cdot \left[ 1 + \frac{(x-\mu)}{\left( \mu^{\frac{2}{3}} - 2(x-\mu) - 1 \right)^{\frac{3}{2}}} \right], \qquad (15)$$

- The expression (8) for coordinate $y$ is given via coordinates $\{x, z\}$, true anomaly $f$, and additional parameter $\alpha$ in (10) ($\alpha = 1$):

$$y \cong \pm\sqrt{\frac{\mu^{\frac{2}{3}}}{\left( \frac{\ddot{z}}{z} \cdot (1 + e \cdot \cos f) + e \cdot \cos f \right)^{\frac{2}{3}}} - \left( (x-\mu+1)^2 + z^2 \right)},$$

$$\frac{\mu^{\frac{2}{3}}}{\left( \frac{\ddot{z}}{z} \cdot (1 + e \cdot \cos f) + e \cdot \cos f \right)^{\frac{2}{3}}} - z^2 \geq (x-\mu+1)^2,$$

- expression for $z$ is given in (12) ($\{C_1, C_2\} = const$, $\alpha = 1$):

$$z = C_1 \cos f + C_2 \sin f$$

The aforementioned *non-linear* ODE of the second order for function $x(f)$ should be solved under the optimizing condition $r_2 \to R_p$, as soon as coordinates $\{x, y, z\}$ are already calculated (while the magnitude of this function $r_2$ should exceed the minimal distance or Roche limit [4-5] for the planet with mass $m_{planet} = \mu$).

Let us also consider the case $\alpha(f) \neq$ const in Eqns. (10)-(11) (which are proved to be valid for the *trapped motion* (7) of infinitesimal mass $m$ in the ER3BP (1)-(4)).

Using (10), we obtain

$$\ddot{z} + \alpha(f) \cdot z = 0, \tag{16}$$

where Eqn. (16) could be transformed by the change of variables $\frac{\dot{z}}{z}$ to the *Riccati* ODE of the first order [10] (in case $z \neq 0$).

Then let us simplify the second equation of system (4) by using the left part of (10)

$$(\ddot{y} + 2\dot{x}) \cdot (1 + e \cdot \cos f) - y = -\frac{y}{|\vec{r}_2|^3} \cdot \left\{ (1-\mu) \cdot \frac{|\vec{r}_2|^3}{|\vec{r}_1|^3} + \mu \right\},$$

$$\left\{ \left( \frac{|\vec{r}_2|}{|\vec{r}_1|} \right)^3 \to 0 \right\} \quad \Rightarrow \quad \frac{1}{|\vec{r}_2|^3} \cong \frac{1}{\mu} \cdot \big(1 + (\alpha - 1) \cdot (1 + e \cdot \cos f)\big) \tag{17}$$

where (in case $z \neq 0$)

$$\alpha(f) = -\frac{\ddot{z}}{z}$$

Furthermore, basing on (16)-(17), we can write out the key *non-linear* ordinary differential equation of the second order in regard to the coordinate $x(f)$ in case of the given function (16) for the coordinate $z$

$$\ddot{x} - 2\dot{y} + (\alpha - 1)\cdot(x-\mu+1) = -\frac{(1-\mu)}{1+e\cdot\cos f}\cdot\left[1 + \frac{(x-\mu)}{\left((x-\mu)^2 + y^2 + z^2\right)^{\frac{3}{2}}}\right], \qquad \Rightarrow$$

$$\left\{ y \cong \pm\sqrt{\frac{\mu^{\frac{2}{3}}}{\left(\alpha(f)\cdot(1+e\cdot\cos f) - e\cdot\cos f\right)^{\frac{2}{3}}} - \left( (x-\mu+1)^2 + z^2 \right)} \right\} \qquad \Rightarrow$$

$$\ddot{x} - 2\dot{y} + (\alpha-1)\cdot(x-\mu+1) = -\frac{(1-\mu)}{1+e\cdot\cos f}\cdot\left[1+\frac{(x-\mu)}{\left(\frac{\mu^{\frac{2}{3}}}{\left(\alpha(f)\cdot(1+e\cdot\cos f) - e\cdot\cos f\right)^{\frac{2}{3}}} - 2(x-\mu) - 1\right)^{\frac{3}{2}}}\right], \quad (18)$$

where expression (8) for coordinate $y$ in (18) is given via coordinates $\{x, z\}$, true anomaly $f$, and the additional parameter $\alpha$ in (10) or (16):

$$y \cong \pm\sqrt{\frac{\mu^{\frac{2}{3}}}{\left(\alpha(f)\cdot(1+e\cdot\cos f) - e\cdot\cos f\right)^{\frac{2}{3}}} - \left( (x-\mu+1)^2 + z^2 \right)},$$

$$\frac{\mu^{\frac{2}{3}}}{\left(\alpha\cdot(1+e\cdot\cos f) - e\cdot\cos f\right)^{\frac{2}{3}}} - z^2 > (x-\mu+1)^2 .$$

The aforementioned *non-linear* ODE of the second order for function $x(f)$ should be solved under the optimizing condition $r_2 \to R_P$, as soon as coordinates $\{x, y, z\}$ are already calculated (while the magnitude of this function $r_2$ should exceed the minimal distances within the Roche-limit [4-5] for the planet with mass $m_{planet}$).

As for the numerical checking of the proper solutions of Eqn. (18), we have tested the cases of Earth, Mars and Venus (see their appropriate parameteres at Table 1 in **Appendix, A1**). By the way, for the Earth it means $z = 0$ in (16)-(18) where we can choose function $\alpha(f)$ absolutely arbitrary, but taking into account that it should be the slowly varying function, e.g. as below

$$\alpha(f) \cong 1000 - 100 \cdot \left( \frac{1 - \exp(\sin f)}{1 + \exp(\sin f)} \right) \tag{19}$$

Meanwhile, this choice of function $\alpha(f)$ (19) can also be applied for calculations in cases of Mars and Venus as well.

We should note that we have used for calculating the data (Table 2-4) the Runge–Kutta fourth-order method with step 0.001 at initial values for Eqns. (16) and (18) as follows: 1) $x_0$ = - 1.0103652222598 and $(\dot{x})_0$ = 0 (for Earth; we consider $z$ = 0); 2) $x_0$ = - 1, $(\dot{x})_0$ = 0, $z_0$ = 0, $(\dot{z})_0$ = - 0.1 (for Mars); 3) $x_0$ = - 1, $(\dot{x})_0$ = 0, $z_0$ = 0, $(\dot{z})_0$ = - 0.3 (for Venus). All the results of numerical calculations for Eqns. (16)-(19) (see Tables 2-4 in **Appendix, A2**) we schematically imagine at Figs.1-5.

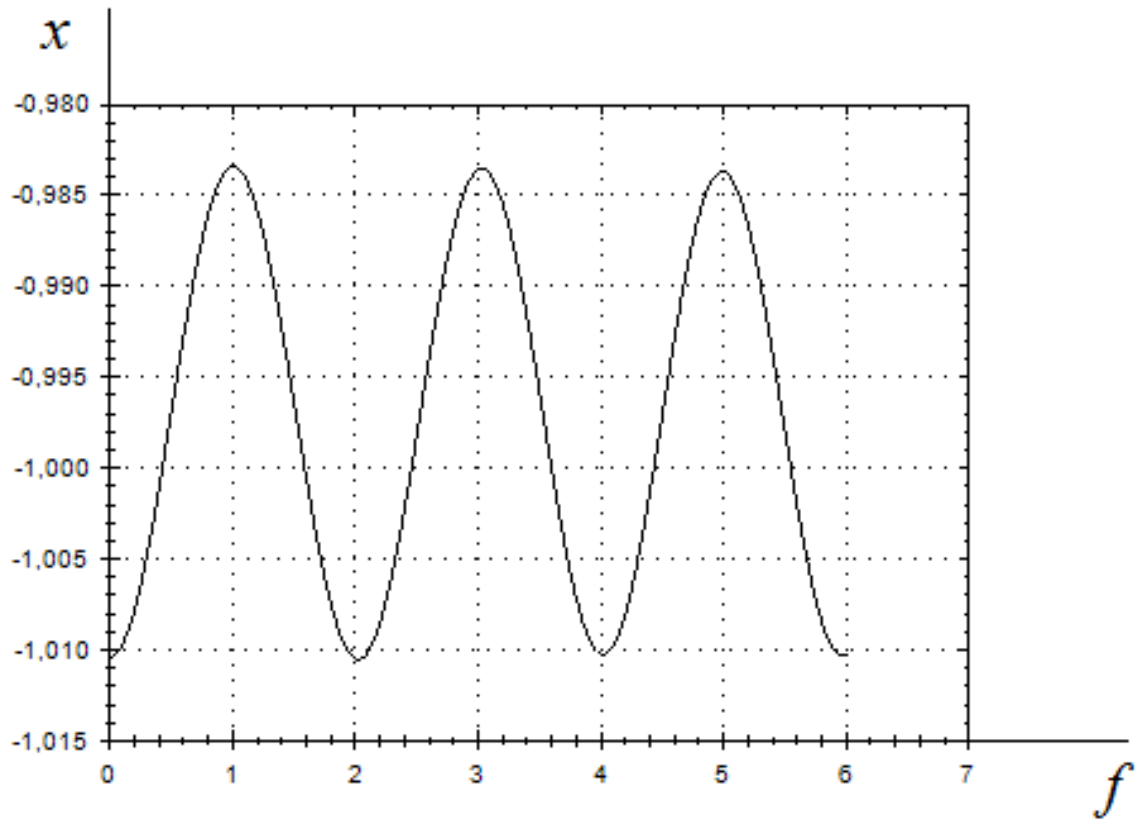

Fig.1. Results of numerical calculations of the coordinate $x$ by Eqn. (18) for Earth.

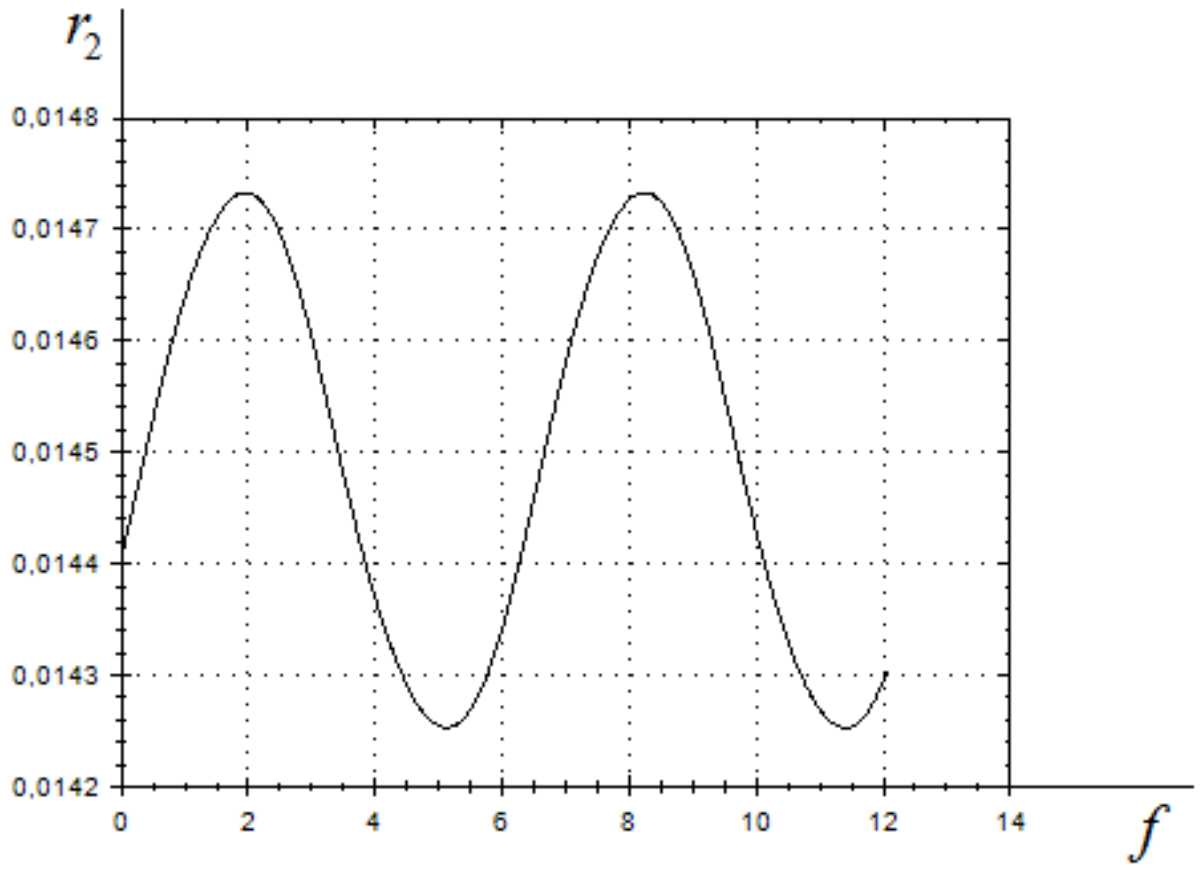

Fig.2. Results of numerical calculations of $r_2$ by using (17) and (19) for Earth.

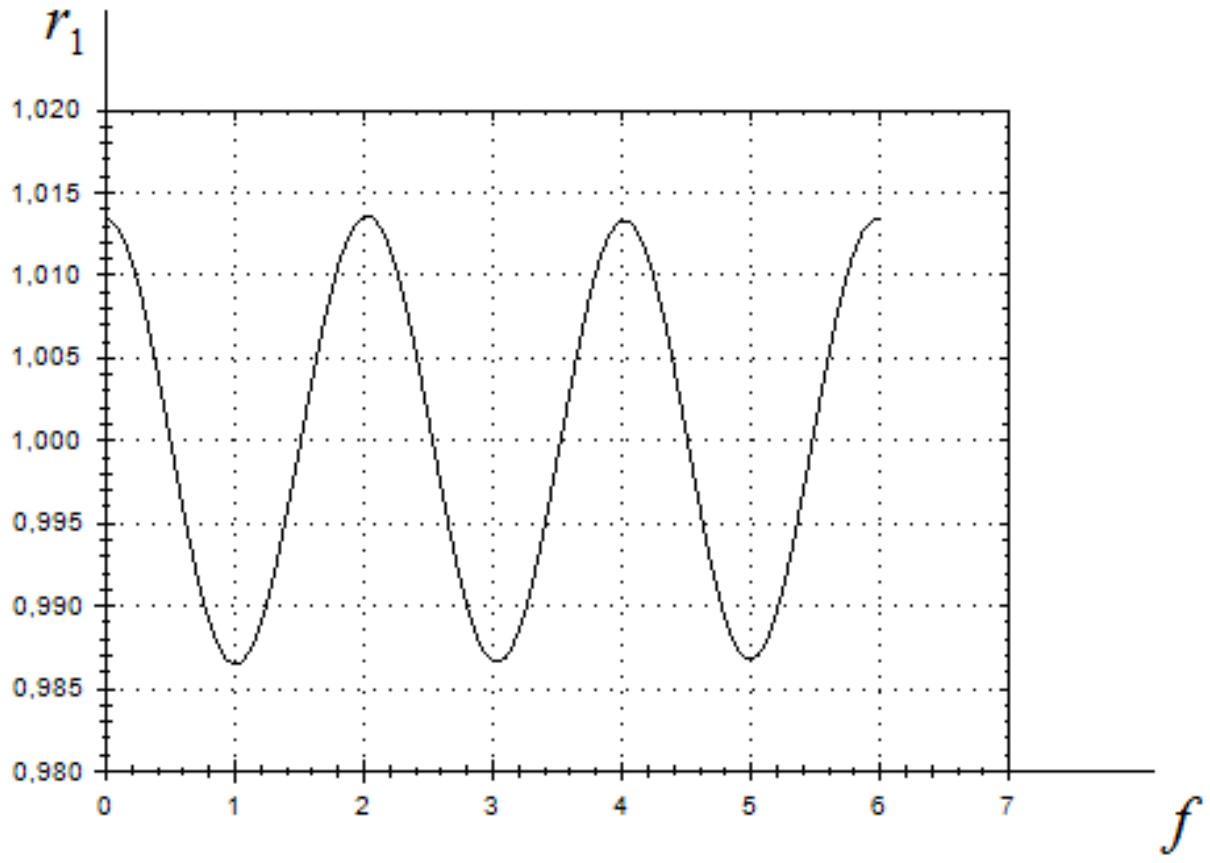

Fig.3. Results of numerical calculations of $r_1$ by Eqns. (17)-(19) for Earth.

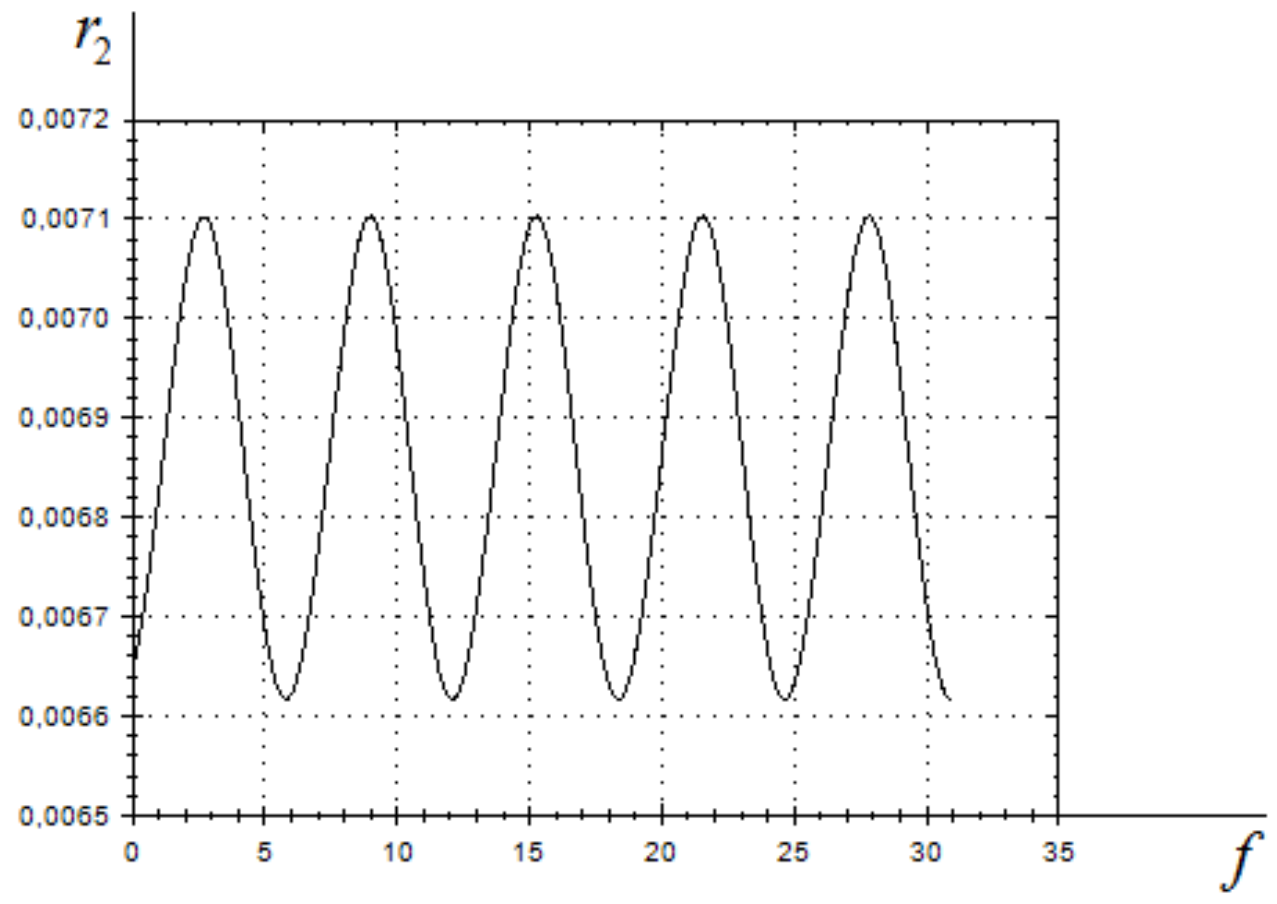

Fig.4. Results of numerical calculations of $r_2$ by Eqns. (16)-(19) for Mars.

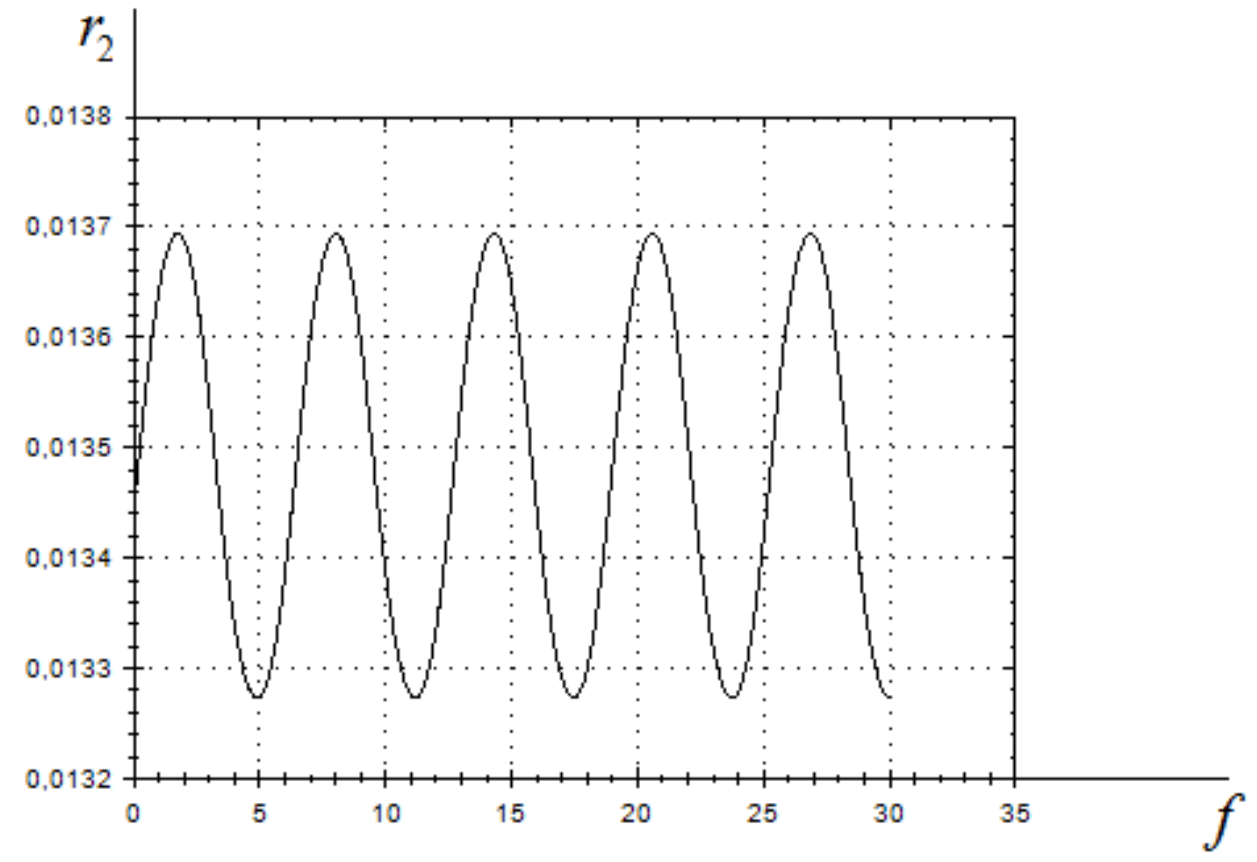

Fig.5. Results of numerical calculations of $r_2$ by Eqns. (16)-(19) for Venus.

## 6. Discussion.

As we can see from the derivation above, equations of motion (4) even for the case of *trapped motion* $\vec{r} = \{x, y, z\}$ (in the sense of additional assumption (7) for the infinitesimal mass $m$ (satellite), which is moving near the planet $m_{planet}$) are proved to be very hard to solve analytically.

Nevertheless, at first step we have succeeded in obtaining the elegant expression for the differential invariant (6), which interconnects the second and third equations of system (4). The aforesaid invariant (6) yields the equation (9) of a *Riccati*-type for the coordinate $z$ which should determine the proper solution (for coordinate $z$) if the solutions for coordinates $\{x, y\}$ are already obtained. Then we suggest a kind of reduction in a form (10) for the Eqn. (9) which let us obtain the classical analytical (*quasi-periodic*) solution (12) for coordinate $z$ (if additional function $\alpha$ is considered to be slowly varying parameter or *circa constant*), as well as it let us obtain in equation (8) the *approximated* solution for coordinate $y$, which was derived under assumption (7). Hereafter we have considered the case $\alpha = 1$ for the sake of simplicity. The key *non-linear* ODE in regard to the coordinate $x(f)$ in case of the given function (12) for the coordinate $z$, $\alpha = 1$ (whereas true anomaly $f$ is to be slowly varying independent coordinate) is obtained to be presented in (15).

Meanwhile, the key *non-linear* ODE in regard to coordinate $x(f)$ in case of the given function for coordinate $z$ via (10) is obtained to be presented in a form (18) for $\alpha \neq const$. This result outlines the novelty of the paper which should be discussed accordingly. Indeed, results (16)-(18) and (20) in **Appendix A1** are obtained as generalization of the simple case $\alpha = 1$ which was investigated first in work [1] by F.Cabral (the well known Clohessy-Wiltshire equations for relative motion when $e \neq 0$) but without obtaining expression for $y$ presented by (8) here.

According to our knowledge, semi-analytical solution, obtained by F.Cabral in [1] for simple case $\alpha = 1$ is most close to our general case of solutions, presented by formulae (16)-(18) for $\alpha \neq const$; moreover, such nontrivial type of solution has not been suggested until the current research and the only way to get some kind of information about the intrinsic properties and behavior of even the particular dynamical system (1) or (4) was to calculate approximated solutions by using various numerical methods. So, any new theoretical method or semi-analytical approach for even the particular solving of such the system of equations would be useful on the level of practical applications.

We also determine the *analytical* expression which interrelates the coordinate $x$, depending on the given coordinate $z$ and true anomaly $f$, relative to the additional function $\delta(f)$ in (14), which determines deviation of distance of infinitesimal mass $m$ (to $M_{Sun}$) from variable relative distance $\rho = \dfrac{a_p}{1 + e \cdot \cos f}$ between primaries $M_{Sun}$ and $m_{planet}$.

Ending discussion, let us note also that natural restriction $r_2 \to R_p$ should be valid for the *trapped motion* $\vec{r} = \{x,\ y,\ z\}$ with respect to the aforementioned function $r_2$ in case of the given function for coordinate $z$ in (10) ($R_p$ is the radius of planet $m_{planet}$). It means that we should choose among the solutions of *non-linear* ODE of second order in regard to the coordinate $x(f)$ only the optimized solutions, for which the condition $r_2 \to R_p$ is valid accordingly (while its magnitude should exceed the level of minimal distances outside the Roche-lobe's region [4-5] for the planet with mass $m_{planet}$).

## 7. Conclusion.

In this paper, we present a new ansatz for solving equations of motion for the *trapped orbits* of the infinitesimal mass $m$ (satellite), which is locked in the space trap to be moving near the planet $m_{planet}$ in case of the elliptic restricted problem of three bodies, ER3BP (with Keplerian *elliptic* trajectories of the primaries $M_{Sun}$ and $m_{planet}$ around each other): a new type of the solving procedure is implemented here to obtain the coordinates $\vec{r} = \{x,\ y,\ z\}$ of the infinitesimal mass $m$ with its orbit located near the planet $m_{planet}$. Meanwhile, the system of equations of motion has been successfully explored with respect to the existence of analytical or semi-analytical (approximated) way for presentation of the solution.

We obtain as follows: 1) equation for coordinate $x$ is given via coordinate $y$, true anomaly $f$, and the additional function $\alpha$, which determines the *quasi-periodic* (a *Riccati*-type) character of solution for coordinate $z$, 2) expression for coordinate $y$ is given via coordinate $x$, true anomaly $f$, and the aforementioned parameter $\alpha$, 3) coordinate $z$ is to be *quasi-periodically* varying with respect to the true anomaly $f$.

We have pointed out the optimizing procedure for the *non-linear* ordinary differential equation of second order in regard to the aforementioned coordinate $x(f)$ in case of the given function for coordinate $z$ (which is valid only for the optimized solutions $r_2 \to R_p$, while the magnitude of function $r_2$ should exceed the level of minimal distances outside the Roche-limit for the planet).

The suggested approach can be used in future researches for optimizing the maneuvers of spacecraft which is moving near the planet $m_{planet} = \mu$ in case of the elliptic restricted problem of three bodies (ER3BP).

So, in this case we also could suggest a scheme for $r_2$-optimizing for the maneuvers of spacecraft which is moving near the planet $m_{planet}$.

Also, some remarkable articles should be cited, which concern the problem under consideration, [12-24] and [25-33]; the results of the most remarkable and comprehensive works (in the sense of algorithms of obtaining the families of solutions in ER3BP) should be outlined and commented additionally hereto. In [17] Dr. J.Singh and A.Umar describe the motion around the collinear libration points in the elliptic R3BP with a bigger triaxial primary. In work [19] E.I.Abouelmagd and M.A.Sharaf classified trajectories of test particle around the libration points in the restricted three-body problem with the effect of radiation and oblateness. In [21], B.S.Kushvah, *et al.* explored the stability (as authors say, of nonlinear character) in the generalised photogravitational restricted three body problem with additional influence of Poynting-Robertson drag. In [28], P.Wiegert, *et al.* investigated the problem of stability of quasi satellites in the outer solar system. In the profound work of W.E. Wiesel [29], a lot of theoretical and numerical findings regarding stable orbits around of the martian moons was established thoroughsly (he e.g. found that stable retro-grade orbits exist about both moons, staying in the moon vicinity for at least 25 days, and quite probably longer).

## 8. Acknowledgements.

Sergey Ershkov appreciates advice of Dr. V.V. Sidorenko during process of meaningful navigation through special literature on celestial mechanics regarding the subject of research. Authors are thankful to unknown esteemed Reviewers with respect to their valuable efforts and advices which have improved structure of the article significantly.

**Appendix, A1 (estimation of possible orbits according to (15)).**

Let us we estimate possible orbits for moons in Solar system [15] which are in agreement with condition reported in (15), $\left| \vec{r}_2 \right| \cong \mu^{\frac{1}{3}}$ :

Table 1. Comparison of the averaged parameters of the moons in Solar system.

| Masses of the planets (*Solar system*), kg | Ratio $\left( \frac{m_{planet} / m_{Earth}}{M_{Sun} / m_{Earth}} \right) = \mu$ | Semimajor axis $a_p$ of the planet, AU | Possible distance $\left\| \vec{r}_2 \right\| \cong a_p \cdot \left( \frac{\mu}{3} \right)^{\frac{1}{3}}$ (*between moon-planet*), AU ($10^3$ km) | Distance of real moon (*between Moon-Planet*), $10^3$ km |
|---|---|---|---|---|
| Mercury, $3.3 \cdot 10^{23}$ | $\left( \frac{0.055}{332{,}946} \right) = 0.165 \cdot 10^{-6}$ | 0.387 AU | 0.015 AU (2,208) | |
| Venus, $4.87 \cdot 10^{24}$ | $\left( \frac{0.815}{332{,}946} \right) = 2.448 \cdot 10^{-6}$ | 0.723 AU | 0.068 AU (10,124) | |
| Earth, $5.97 \cdot 10^{24}$ + Moon, $7.36 \cdot 10^{22}$ | $\left(\frac{1.0123}{332{,}946}\right) = 3.040 \cdot 10^{-6}$ | 1 AU | 0.101 AU (15,040) | 383.4 |
| Mars, $6.42 \cdot 10^{23}$ | $\left(\frac{0.107}{332{,}946}\right) = 0.321 \cdot 10^{-6}$ | 1.523 AU | 0.073 AU (10,900) | 1) Phobos 9.38<br>2) Deimos 23.46 |
| Jupiter, $1.9 \cdot 10^{27}$ | $\left(\frac{317.8}{332{,}946}\right) = 954.509 \cdot 10^{-6}$ | 5.205 AU | 3.555 AU (531,781) | 1) Ganymede 1,070<br>2) Callisto 1,883<br>3) Io 422<br>4) Europa 671 |
| Saturn, $5.69 \cdot 10^{26}$ | $\left(\frac{95.16}{332{,}946}\right) = 285.812 \cdot 10^{-6}$ | 9.582 | 4.378 (654,964) | 1) Titan 1,222<br>2) Rhea 527<br>3) Iapetus 3,561<br>4) Dione |
| | | | | 377<br>5) Tethys<br>294.6<br>6) Enceladus<br>238 |
| Uranus,<br>$8.69{\cdot}10^{25}$ | $\left(\frac{14.37}{332{,}946}\right) = 43.160 \cdot 10^{-6}$ | 19.201 | 4.673<br>(699,048) | 1) Titania<br>436<br>2) Oberon<br>584<br>3) Ariel:<br>191<br>4) Umbriel:<br>266.3<br>5) Miranda:<br>129.4 |
| Neptune,<br>$1.02{\cdot}10^{26}$ | $\left(\frac{17.15}{332{,}946}\right) = 51.510 \cdot 10^{-6}$ | 30.048 | 7.750<br>(1,159,403) | 1) Triton<br>355<br>2) Proteus<br>118<br>3) Nereid<br>5,513 |
| Pluto,<br>$1.3{\cdot}10^{22}$ | $\left(\frac{0.002}{332{,}946}\right) = 0.006 \cdot 10^{-6}$ | 39.238 | 0.490<br>(73,258) | Charon<br>20 |

Data and results, shown in Table 1, can be compared with previous research, ref. [15] (see Table 1).

As we can see from Table 1, most realistic data (for the problem under consideration) appears to be associated with cases of Mercury or Venus: these planets might have had the moons, which were rotating on circular orbits around their planets. But numerical solving procedure for Eqn. (15) reveals that the ER3BP (in case of Mercury or Venus) allows the existing of the solution on only the limited range of true anomaly *f*. It means that numerical modeling by means of the ER3BP forbids or exclude the existing of the free-gravitating moon near the Mercury (or Venus), taking into account interactions between "Sun-Planet"-system and the aforementioned moon.

As for the case of Venus, we know that during its closest approach to Earth, mutual distance is ~ 38,000·$10^3$ km (which equals to 0.254 AU); so, we obtain from the appropriate data (see Table 1):

$$0.254 \text{ AU} - |\boldsymbol{r}_{2,\, Venus}| - |\boldsymbol{r}_{2,\, Earth}| \cong 0.085 \text{ AU}$$

but if we take into consideration the mass of possible moon of Venus, the aforementioned difference would tend to zero. So, such (possible) moon of Venus was to interact to the Hill sphere of Earth at their closest approaches to each other, that's why its motion was obviously not to be stable. Meanwhile, we could estimate aforesaid minimal mass of possible moon of Venus, for which the (closest) relative distance 0.085 AU would disappear at all.

Let us also we estimate parameter $\alpha(f)$ in expression (11), taking into account the approximation (7):

$$\alpha = \frac{1}{1+e\cdot\cos f}\cdot\left[e\cdot\cos f + \frac{(1-\mu)}{\left((x-\mu)^2 + y^2 + (z)^2\right)^{\frac{3}{2}}} + \frac{\mu}{\left((x-\mu+1)^2 + y^2 + (z)^2\right)^{\frac{3}{2}}}\right] \Rightarrow$$

$$\alpha = \frac{1}{1+e\cdot\cos f}\cdot\left[e\cdot\cos f + \frac{1}{\left((x-\mu+1)^2 + y^2 + z^2\right)^{\frac{3}{2}}}\cdot\left\{(1-\mu)\cdot\left(\frac{\left((x-\mu+1)^2 + y^2 + z^2\right)^{\frac{1}{2}}}{\left((x-\mu)^2 + y^2 + z^2\right)^{\frac{1}{2}}}\right)^3 + \mu\right\}\right] \Rightarrow$$

$$\alpha \cong \frac{1}{1+e\cdot\cos f}\cdot\left[e\cdot\cos f + \frac{1}{r_2^{\;3}}\cdot\{\mu\}\right] \qquad (20)$$

Such the result (20) for estimating the parameter $\alpha(f)$ in expression (11) can be compared with the result of F.Cabral for the simple case $\alpha$ = 1 which was investigated first in work [1].

Obviously, the case of function $\alpha$ to be the slowly varying parameter or *circa constant* corresponds to the condition $r_2 \cong$ const << $r_1$ (where $(r_2)^3 \cong \mu$, but $r_1 \cong 1$). For example, if even we choose $r_2$ = 0.05, $\mu$ = 0.001, $e$ = 0.015, estimations according (20) should yield as follows

$$\alpha \cong \frac{1}{1+(0.015)\cdot\cos f}\cdot\left[(0.015)\cdot\cos f\ +\ 8\right] \ \Rightarrow\ \alpha_{\min} = 7.867\,,\quad \alpha_{\max} = 8.137\quad (8 \pm 1.7\%)$$

**Appendix, A2 (results of numerical calculations for Eqns. (16)-(19)).**

Let us present all the results of numerical calculations for Eqn. (16)-(19) below:

Table 2. Results of numerical calculations for Eqn. (16)-(19) (case of Earth).

| $f$ , rad | $x$ | $y$ | $r_1$ | $r_2$ | $\alpha$ | $\delta$ |
|---|---|---|---|---|---|---|
| 0 | -1.01037 | 0 | 1.013419 | 0.014405 | 1000 | 0.030135 |
| 0.1 | -0.98354 | 0.005301 | 0.986595 | 0.014430 | 995.0125 | 0.003228 |
| 0.2 | -1.01040 | 0.005327 | 1.013451 | 0.014455 | 990.0991 | 0.029839 |
| 0.3 | -0.98351 | 0.005354 | 0.986562 | 0.014480 | 985.3306 | 0.002543 |
| 0.4 | -1.01043 | 0.005394 | 1.013485 | 0.014505 | 980.7715 | 0.028902 |
| 0.5 | -0.98348 | 0.005438 | 0.986530 | 0.014530 | 976.4776 | 0.001230 |
| 0.6 | -1.01045 | 0.005525 | 1.013503 | 0.014554 | 972.4948 | 0.027339 |
| 0.7 | -0.98348 | 0.005623 | 0.986533 | 0.014577 | 968.8588 | -0.000630 |
| 0.8 | -1.01041 | 0.005810 | 1.013461 | 0.014599 | 965.5951 | 0.025166 |
| 0.9 | -0.98357 | 0.006008 | 0.986623 | 0.014620 | 962.7207 | -0.002920 |
| 1.0 | -1.01025 | 0.006333 | 1.013304 | 0.014639 | 960.245 | 0.022406 |
| 1.1 | -0.98379 | 0.006652 | 0.986856 | 0.014657 | 958.1721 | -0.00549 |
| 1.2 | -1.00992 | 0.007111 | 1.012986 | 0.014673 | 956.5022 | 0.019109 |
| 1.3 | -0.98420 | 0.007532 | 0.987270 | 0.014687 | 955.2333 | -0.00820 |
| 1.4 | -1.00941 | 0.008082 | 1.012480 | 0.014700 | 954.3624 | 0.015361 |
| 1.5 | -0.98481 | 0.008562 | 0.987882 | 0.014710 | 953.8868 | -0.010920 |
| 1.6 | -1.00871 | 0.009144 | 1.011787 | 0.014719 | 953.8051 | 0.011291 |
| 1.7 | -0.98559 | 0.009632 | 0.988672 | 0.014726 | 954.1167 | -0.013520 |
| 1.8 | -1.00785 | 0.010189 | 1.010940 | 0.014730 | 954.8229 | 0.007063 |
| 1.9 | -0.98650 | 0.010640 | 0.989593 | 0.014733 | 955.9259 | -0.015930 |
| 2.0 | -1.00690 | 0.011130 | 1.010000 | 0.014733 | 957.4287 | 0.002875 |
| 2.1 | -0.98746 | 0.011510 | 0.990568 | 0.014730 | 959.3339 | -0.018090 |
| 2.2 | -1.00594 | 0.011905 | 1.009051 | 0.014726 | 961.6422 | -0.001050 |
| 2.3 | -0.98839 | 0.012196 | 0.991503 | 0.014718 | 964.3516 | -0.019950 |
| 2.4 | -1.00508 | 0.012485 | 1.008189 | 0.014709 | 967.4549 | -0.004510 |
| 2.5 | -0.98918 | 0.012679 | 0.992299 | 0.014696 | 970.9387 | -0.021510 |
| 2.6 | -1.00439 | 0.012864 | 1.007510 | 0.014682 | 974.781 | -0.007270 |
| 2.7 | -0.98974 | 0.012962 | 0.992863 | 0.014665 | 978.9504 | -0.022750 |
| 2.8 | -1.00398 | 0.013047 | 1.007098 | 0.014646 | 983.4055 | -0.009180 |
| 2.9 | -0.99000 | 0.013053 | 0.993120 | 0.014625 | 988.0943 | -0.023660 |
| 3.0 | -1.00389 | 0.013043 | 1.007013 | 0.014603 | 992.9557 | -0.010110 |
| 3.1 | -0.98991 | 0.012954 | 0.993028 | 0.014579 | 997.9213 | -0.024250 |
| 3.2 | -1.00416 | 0.012848 | 1.007279 | 0.014555 | 1002.918 | -0.009990 |
| 3.3 | -0.98946 | 0.012654 | 0.99258 | 0.014530 | 1007.871 | -0.024490 |
| 3.4 | -1.00477 | 0.012441 | 1.007881 | 0.014505 | 1012.708 | -0.008830 |
| 3.5 | -0.98870 | 0.012125 | 0.991816 | 0.014481 | 1017.361 | -0.024360 |
| 3.6 | -1.00565 | 0.011790 | 1.008759 | 0.014456 | 1021.772 | -0.006720 |
| 3.7 | -0.98771 | 0.011336 | 0.990816 | 0.014433 | 1025.889 | -0.023810 |
| 3.8 | -1.00672 | 0.010869 | 1.009813 | 0.014411 | 1029.673 | -0.003820 |
| 3.9 | -0.98661 | 0.010270 | 0.989698 | 0.014389 | 1033.094 | -0.022800 |
| 4.0 | -1.00783 | 0.009679 | 1.010915 | 0.014369 | 1036.132 | -0.000320 |
| 4.1 | -0.98552 | 0.008957 | 0.988602 | 0.014351 | 1038.774 | -0.021270 |
| 4.2 | -1.00886 | 0.008281 | 1.011926 | 0.014333 | 1041.015 | 0.003522 |
| 4.3 | -0.98461 | 0.007506 | 0.987672 | 0.014318 | 1042.852 | -0.019190 |
| 4.4 | -1.00965 | 0.006839 | 1.012714 | 0.014303 | 1044.287 | 0.007462 |
| 4.5 | -0.98397 | 0.006161 | 0.987031 | 0.014291 | 1045.324 | -0.016570 |
| 4.6 | -1.01012 | 0.005681 | 1.013175 | 0.014280 | 1045.963 | 0.011264 |
| 4.7 | -0.98371 | 0.005347 | 0.986765 | 0.014271 | 1046.209 | -0.013450 |
| 4.8 | -1.01019 | 0.005278 | 1.013248 | 0.014263 | 1046.061 | 0.014733 |
| 4.9 | -0.98385 | 0.005474 | 0.986907 | 0.014258 | 1045.519 | -0.009930 |
| 5.0 | -1.00987 | 0.005847 | 1.012927 | 0.014254 | 1044.581 | 0.017726 |
| 5.1 | -0.98437 | 0.006440 | 0.987436 | 0.014253 | 1043.245 | -0.006180 |
| 5.2 | -1.00919 | 0.007043 | 1.012259 | 0.014254 | 1041.508 | 0.020161 |
| 5.3 | -0.98520 | 0.007771 | 0.988272 | 0.014256 | 1039.367 | -0.002390 |
| 5.4 | -1.00826 | 0.008407 | 1.011339 | 0.014262 | 1036.824 | 0.022013 |
| 5.5 | -0.98622 | 0.009101 | 0.989299 | 0.014269 | 1033.883 | 0.001203 |
| 5.6 | -1.00721 | 0.009663 | 1.010294 | 0.014279 | 1030.555 | 0.023307 |
| 5.7 | -0.98728 | 0.010244 | 0.990377 | 0.014292 | 1026.859 | 0.004369 |
| 5.8 | -1.00617 | 0.010688 | 1.009266 | 0.014306 | 1022.821 | 0.024096 |
| 5.9 | -0.98826 | 0.011129 | 0.991365 | 0.014323 | 1018.479 | 0.006888 |
| 6.0 | -1.00528 | 0.011443 | 1.008386 | 0.014342 | 1013.881 | 0.024446 |
| 6.1 | -0.98903 | 0.011746 | 0.992142 | 0.014363 | 1009.083 | 0.008583 |
| 6.2 | -1.00465 | 0.011935 | 1.007760 | 0.014386 | 1004.152 | 0.024419 |
| 6.3 | -0.98950 | 0.012113 | 0.992621 | 0.014409 | 999.1593 | 0.009334 |
| 6.4 | -1.00434 | 0.012189 | 1.007456 | 0.014434 | 994.1791 | 0.024060 |
| 6.5 | -0.98964 | 0.012255 | 0.992755 | 0.014459 | 989.2853 | 0.009086 |
| 6.6 | -1.00438 | 0.012224 | 1.007497 | 0.014484 | 984.5477 | 0.023394 |
| 6.7 | -0.98943 | 0.012185 | 0.992546 | 0.014509 | 980.0295 | 0.007853 |
| 6.8 | -1.00475 | 0.012048 | 1.007863 | 0.014534 | 975.7850 | 0.022427 |
| 6.9 | -0.98892 | 0.011907 | 0.992032 | 0.014558 | 971.8583 | 0.005710 |

Table 3. Results of numerical calculations for Eqn. (16)-(19) (case of Mars).

| $f$, rad | $x$ | $y$ | $z$ | $r_1$ | $r_2$ | $\alpha$ | $\delta$ |
|---|---|---|---|---|---|---|---|
| 0 | -1 | 0 | 0 | 1.000343 | 0.006645 | 1000 | 0.086266 |
| 0.1 | -0.99936 | 0.006649 | 0.000053 | 0.999701 | 0.006657 | 995.0125 | 0.085231 |
| 0.2 | -1 | 0.006663 | -8.1E-05 | 1.000344 | 0.006671 | 990.0991 | 0.084699 |
| 0.3 | -0.99936 | 0.006679 | 0.000085 | 0.9997 | 0.006687 | 985.3306 | 0.082102 |
| 0.4 | -1 | 0.006696 | -6.5E-05 | 1.000345 | 0.006704 | 980.7715 | 0.080026 |
| 0.5 | -0.99936 | 0.006713 | 0.000023 | 0.9997 | 0.006722 | 976.4776 | 0.075906 |
| 0.6 | -1 | 0.006732 | 0.000041 | 1.000346 | 0.006741 | 972.4948 | 0.072342 |
| 0.7 | -0.99936 | 0.00675 | -0.00012 | 0.9997 | 0.006761 | 968.8588 | 0.066773 |
| 0.8 | -1 | 0.006767 | 0.000225 | 1.000345 | 0.006783 | 965.5951 | 0.06181 |
| 0.9 | -0.99936 | 0.006783 | -0.00034 | 0.999703 | 0.006804 | 962.7207 | 0.054909 |
| 1 | -1 | 0.006795 | 0.000471 | 1.000341 | 0.006827 | 960.245 | 0.048675 |
| 1.1 | -0.99937 | 0.006805 | -0.00061 | 0.99971 | 0.006849 | 958.1721 | 0.040604 |
| 1.2 | -0.99999 | 0.006812 | 0.00076 | 1.000334 | 0.006872 | 956.5022 | 0.033274 |
| 1.3 | -0.99938 | 0.006814 | -0.00092 | 0.99972 | 0.006895 | 955.2333 | 0.024249 |
| 1.4 | -0.99998 | 0.006813 | 0.001072 | 1.000322 | 0.006918 | 954.3624 | 0.016048 |
| 1.5 | -0.99939 | 0.006807 | -0.00123 | 0.999736 | 0.00694 | 953.8868 | 0.006341 |
| 1.6 | -0.99996 | 0.006799 | 0.001385 | 1.000306 | 0.006962 | 953.8051 | -0.00245 |
| 1.7 | -0.99941 | 0.006787 | -0.00154 | 0.999755 | 0.006983 | 954.1167 | -0.01251 |
| 1.8 | -0.99994 | 0.006773 | 0.00168 | 1.000285 | 0.007004 | 954.8229 | -0.02154 |
| 1.9 | -0.99943 | 0.006757 | -0.00182 | 0.999778 | 0.007022 | 955.9259 | -0.03156 |
| 2 | -0.99992 | 0.00674 | 0.001941 | 1.000263 | 0.00704 | 957.4287 | -0.04045 |
| 2.1 | -0.99946 | 0.006723 | -0.00205 | 0.999801 | 0.007056 | 959.3339 | -0.05002 |
| 2.2 | -0.99989 | 0.006706 | 0.002155 | 1.00024 | 0.00707 | 961.6422 | -0.05832 |
| 2.3 | -0.99948 | 0.00669 | -0.00224 | 0.999824 | 0.007081 | 964.3516 | -0.06699 |
| 2.4 | -0.99987 | 0.006675 | 0.002316 | 1.000219 | 0.007091 | 967.4549 | -0.07426 |
| 2.5 | -0.9995 | 0.006662 | -0.00238 | 0.999843 | 0.007097 | 970.9387 | -0.0816 |
| 2.6 | -0.99986 | 0.006651 | 0.00242 | 1.000202 | 0.007101 | 974.781 | -0.0874 |
| 2.7 | -0.99951 | 0.006641 | -0.00245 | 0.999857 | 0.007103 | 978.9504 | -0.09302 |
| 2.8 | -0.99985 | 0.006634 | 0.002467 | 1.000192 | 0.007101 | 983.4055 | -0.09698 |
| 2.9 | -0.99952 | 0.006629 | -0.00247 | 0.999863 | 0.007097 | 988.0943 | -0.10057 |
| 3 | -0.99984 | 0.006627 | 0.002453 | 1.00019 | 0.00709 | 992.9557 | -0.10242 |
| 3.1 | -0.99952 | 0.006628 | -0.00242 | 0.999861 | 0.007081 | 997.9213 | -0.10379 |
| 3.2 | -0.99985 | 0.006632 | 0.002376 | 1.000196 | 0.007069 | 1002.918 | -0.10336 |
| 3.3 | -0.9995 | 0.006639 | -0.00231 | 0.99985 | 0.007054 | 1007.871 | -0.10247 |
| 3.4 | -0.99987 | 0.00665 | 0.002228 | 1.00021 | 0.007038 | 1012.708 | -0.09975 |
| 3.5 | -0.99949 | 0.006665 | -0.00212 | 0.999832 | 0.00702 | 1017.361 | -0.09669 |
| 3.6 | -0.99989 | 0.006683 | 0.001999 | 1.000231 | 0.007 | 1021.772 | -0.09182 |
| 3.7 | -0.99946 | 0.006703 | -0.00185 | 0.999808 | 0.006979 | 1025.889 | -0.08682 |
| 3.8 | -0.99991 | 0.006725 | 0.001684 | 1.000257 | 0.006957 | 1029.673 | -0.08007 |
| 3.9 | -0.99944 | 0.006748 | -0.00149 | 0.99978 | 0.006934 | 1033.094 | -0.07346 |
| 4 | -0.99994 | 0.006769 | 0.001282 | 1.000283 | 0.006911 | 1036.132 | -0.06518 |
| 4.1 | -0.99941 | 0.006786 | -0.00105 | 0.999753 | 0.006887 | 1038.774 | -0.05737 |
| 4.2 | -0.99996 | 0.006798 | 0.000801 | 1.000307 | 0.006863 | 1041.015 | -0.048 |
| 4.3 | -0.99939 | 0.006803 | -0.00054 | 0.99973 | 0.006839 | 1042.852 | -0.03942 |
| 4.4 | -0.99998 | 0.006799 | 0.000261 | 1.000326 | 0.006815 | 1044.287 | -0.02942 |
| 4.5 | -0.99937 | 0.006785 | 0.000023 | 0.999713 | 0.006792 | 1045.324 | -0.0205 |
| 4.6 | -0.99999 | 0.006759 | -0.00031 | 1.000337 | 0.006769 | 1045.963 | -0.01032 |
| 4.7 | -0.99936 | 0.006722 | 0.000598 | 0.999705 | 0.006747 | 1046.209 | -0.00146 |
| 4.8 | -1 | 0.006674 | -0.00088 | 1.000339 | 0.006727 | 1046.061 | 0.008497 |
| 4.9 | -0.99936 | 0.006616 | 0.001153 | 0.999707 | 0.006707 | 1045.519 | 0.016937 |
| 5 | -0.99999 | 0.00655 | -0.00141 | 1.000331 | 0.006689 | 1044.581 | 0.026303 |
| 5.1 | -0.99938 | 0.006478 | 0.001657 | 0.999718 | 0.006673 | 1043.245 | 0.034029 |
| 5.2 | -0.99997 | 0.006404 | -0.00188 | 1.000315 | 0.006659 | 1041.508 | 0.042498 |
| 5.3 | -0.99939 | 0.006328 | 0.002087 | 0.999737 | 0.006646 | 1039.367 | 0.049267 |
| 5.4 | -0.99995 | 0.006255 | -0.00227 | 1.000293 | 0.006636 | 1036.824 | 0.056595 |
| 5.5 | -0.99942 | 0.006186 | 0.002429 | 0.999761 | 0.006628 | 1033.883 | 0.062215 |
| 5.6 | -0.99993 | 0.006124 | -0.00257 | 1.000268 | 0.006622 | 1030.555 | 0.068218 |
| 5.7 | -0.99944 | 0.006069 | 0.002682 | 0.999786 | 0.006618 | 1026.859 | 0.07254 |
| 5.8 | -0.9999 | 0.006024 | -0.00278 | 1.000244 | 0.006617 | 1022.821 | 0.077086 |
| 5.9 | -0.99947 | 0.005988 | 0.002855 | 0.999809 | 0.006619 | 1018.479 | 0.080001 |
| 6 | -0.99988 | 0.005961 | -0.00292 | 1.000223 | 0.006622 | 1013.881 | 0.083007 |
| 6.1 | -0.99949 | 0.005943 | 0.002964 | 0.999828 | 0.006628 | 1009.083 | 0.084435 |
| 6.2 | -0.99987 | 0.005933 | -0.003 | 1.000208 | 0.006637 | 1004.152 | 0.08586 |

Table 4. Results of numerical calculations for Eqn. (16)-(19) (case of Venus).

| $f$, rad | $x$ | $y$ | $z$ | $r_1$ | $r_2$ | $\alpha$ | $\delta$ |
|---|---|---|---|---|---|---|---|
| 0 | -1 | 0.013221 | 0 | 1.002535 | 0.013446 | 1000 | 0.009487 |
| 0.1 | -0.9951 | 0.013243 | 0.000159 | 0.997637 | 0.013469 | 995.0125 | 0.004554 |
| 0.2 | -1.00001 | 0.013264 | -0.00024 | 1.002541 | 0.013491 | 990.0991 | 0.009355 |
| 0.3 | -0.9951 | 0.013286 | 0.000255 | 0.997631 | 0.013514 | 985.3306 | 0.004274 |
| 0.4 | -1.00001 | 0.013309 | -0.0002 | 1.002548 | 0.013536 | 980.7715 | 0.008954 |
| 0.5 | -0.99509 | 0.013331 | 0.000068 | 0.997627 | 0.013557 | 976.4776 | 0.003733 |
| 0.6 | -1.00002 | 0.013351 | 0.000122 | 1.002552 | 0.013577 | 972.4948 | 0.008296 |
| 0.7 | -0.99509 | 0.013367 | -0.00037 | 0.997631 | 0.013596 | 968.8588 | 0.002956 |
| 0.8 | -1.00001 | 0.013373 | 0.000674 | 1.002544 | 0.013613 | 965.5951 | 0.007397 |
| 0.9 | -0.99511 | 0.01337 | -0.00102 | 0.997651 | 0.013629 | 962.7207 | 0.001983 |
| 1 | -0.99998 | 0.013352 | 0.001413 | 1.002515 | 0.013644 | 960.245 | 0.006283 |
| 1.1 | -0.99516 | 0.013319 | -0.00184 | 0.997698 | 0.013656 | 958.1721 | 0.000863 |
| 1.2 | -0.99992 | 0.013266 | 0.002281 | 1.002456 | 0.013667 | 956.5022 | 0.004987 |
| 1.3 | -0.99524 | 0.013197 | -0.00275 | 0.997777 | 0.013676 | 955.2333 | -0.00035 |
| 1.4 | -0.99982 | 0.013104 | 0.003217 | 1.002363 | 0.013683 | 954.3624 | 0.003551 |
| 1.5 | -0.99535 | 0.012997 | -0.00369 | 0.997892 | 0.013689 | 953.8868 | -0.00161 |
| 1.6 | -0.9997 | 0.012869 | 0.004156 | 1.002234 | 0.013692 | 953.8051 | 0.00203 |
| 1.7 | -0.9955 | 0.012731 | -0.00461 | 0.998039 | 0.013694 | 954.1167 | -0.00286 |
| 1.8 | -0.99954 | 0.012577 | 0.005041 | 1.002077 | 0.013694 | 954.8229 | 0.000484 |
| 1.9 | -0.99567 | 0.01242 | -0.00545 | 0.998209 | 0.013692 | 955.9259 | -0.00406 |
| 2 | -0.99936 | 0.012254 | 0.005822 | 1.001903 | 0.013687 | 957.4287 | -0.00102 |
| 2.1 | -0.99585 | 0.012095 | -0.00616 | 0.998387 | 0.013681 | 959.3339 | -0.00516 |
| 2.2 | -0.99919 | 0.011936 | 0.006465 | 1.001727 | 0.013673 | 961.6422 | -0.00241 |
| 2.3 | -0.99602 | 0.011792 | -0.00673 | 0.998558 | 0.013663 | 964.3516 | -0.00613 |
| 2.4 | -0.99903 | 0.011657 | 0.006948 | 1.001567 | 0.01365 | 967.4549 | -0.00362 |
| 2.5 | -0.99616 | 0.011543 | -0.00713 | 0.998703 | 0.013636 | 970.9387 | -0.00694 |
| 2.6 | -0.9989 | 0.011444 | 0.007261 | 1.001441 | 0.01362 | 974.781 | -0.00459 |
| 2.7 | -0.99627 | 0.011371 | -0.00735 | 0.998805 | 0.013602 | 978.9504 | -0.00756 |
| 2.8 | -0.99883 | 0.011319 | 0.0074 | 1.001364 | 0.013583 | 983.4055 | -0.00528 |
| 2.9 | -0.99631 | 0.011296 | -0.0074 | 0.998851 | 0.013562 | 988.0943 | -0.00799 |
| 3 | -0.99881 | 0.011296 | 0.00736 | 1.001348 | 0.01354 | 992.9557 | -0.00563 |
| 3.1 | -0.9963 | 0.011328 | -0.00727 | 0.998833 | 0.013518 | 997.9213 | -0.00821 |
| 3.2 | -0.99886 | 0.011385 | 0.007128 | 1.001397 | 0.013496 | 1002.918 | -0.00564 |
| 3.3 | -0.99621 | 0.011474 | -0.00693 | 0.998751 | 0.013473 | 1007.871 | -0.00821 |
| 3.4 | -0.99897 | 0.011587 | 0.006683 | 1.001508 | 0.013451 | 1012.708 | -0.00531 |
| 3.5 | -0.99607 | 0.011729 | -0.00637 | 0.998611 | 0.01343 | 1017.361 | -0.00799 |
| 3.6 | -0.99913 | 0.011888 | 0.005998 | 1.001671 | 0.013409 | 1021.772 | -0.00465 |
| 3.7 | -0.99589 | 0.012067 | -0.00556 | 0.998427 | 0.01339 | 1025.889 | -0.00755 |
| 3.8 | -0.99933 | 0.012252 | 0.005053 | 1.001865 | 0.013371 | 1029.673 | -0.0037 |
| 3.9 | -0.99569 | 0.012441 | -0.00448 | 0.998221 | 0.013354 | 1033.094 | -0.00689 |
| 4 | -0.99953 | 0.012618 | 0.003846 | 1.002069 | 0.013339 | 1036.132 | -0.00253 |
| 4.1 | -0.99548 | 0.012781 | -0.00315 | 0.998018 | 0.013325 | 1038.774 | -0.00602 |
| 4.2 | -0.99972 | 0.012914 | 0.002403 | 1.002255 | 0.013313 | 1041.015 | -0.00119 |
| 4.3 | -0.99531 | 0.013013 | -0.00161 | 0.997843 | 0.013303 | 1042.852 | -0.00497 |
| 4.4 | -0.99987 | 0.013067 | 0.000782 | 1.0024 | 0.013294 | 1044.287 | 0.000244 |
| 4.5 | -0.99519 | 0.013074 | 0.00007 | 0.99772 | 0.013286 | 1045.324 | -0.00376 |
| 4.6 | -0.99995 | 0.013029 | -0.00093 | 1.002485 | 0.01328 | 1045.963 | 0.001699 |
| 4.7 | -0.99513 | 0.01293 | 0.001793 | 0.997664 | 0.013276 | 1046.209 | -0.00242 |
| 4.8 | -0.99997 | 0.012783 | -0.00264 | 1.002498 | 0.013274 | 1046.061 | 0.00311 |
| 4.9 | -0.99515 | 0.012587 | 0.003458 | 0.997683 | 0.013273 | 1045.519 | -0.00101 |
| 5 | -0.99991 | 0.012357 | -0.00424 | 1.002439 | 0.013274 | 1044.581 | 0.004421 |
| 5.1 | -0.99524 | 0.012091 | 0.004972 | 0.997772 | 0.013277 | 1043.245 | 0.000411 |
| 5.2 | -0.99978 | 0.011812 | -0.00565 | 1.002317 | 0.013281 | 1041.508 | 0.005586 |
| 5.3 | -0.99538 | 0.011518 | 0.006261 | 0.997919 | 0.013288 | 1039.367 | 0.001784 |
| 5.4 | -0.99962 | 0.011233 | -0.00681 | 1.00215 | 0.013296 | 1036.824 | 0.006573 |
| 5.5 | -0.99557 | 0.010956 | 0.007286 | 0.998102 | 0.013307 | 1033.883 | 0.003038 |
| 5.6 | -0.99943 | 0.010706 | -0.0077 | 1.00196 | 0.013319 | 1030.555 | 0.00736 |
| 5.7 | -0.99576 | 0.01048 | 0.008045 | 0.998296 | 0.013333 | 1026.859 | 0.004105 |
| 5.8 | -0.99924 | 0.010292 | -0.00833 | 1.001774 | 0.013349 | 1022.821 | 0.007935 |
| 5.9 | -0.99594 | 0.010134 | 0.008565 | 0.998474 | 0.013366 | 1018.479 | 0.004925 |
| 6 | -0.99908 | 0.010015 | -0.00875 | 1.001615 | 0.013386 | 1013.881 | 0.008291 |
| 6.1 | -0.99608 | 0.009924 | 0.008892 | 0.998616 | 0.013406 | 1009.083 | 0.005452 |
| 6.2 | -0.99897 | 0.009864 | -0.009 | 1.001502 | 0.013428 | 1004.152 | 0.00843 |

We should note that we have used for calculating the data (Table 2-4) the Runge–Kutta fourth-order method with step 0.001 at initial values for Eqns. (16) and (18) as follows: 1) $x_0$ = - 1.0103652222598 and $(\dot{x})_0$ = 0 (for Earth; we consider $z$ = 0); 2) $x_0$ = - 1, $(\dot{x})_0$ = 0, $z_0$ = 0, $(\dot{z})_0$ = - 0.1 (for Mars); 3) $x_0$ = - 1, $(\dot{x})_0$ = 0, $z_0$ = 0, $(\dot{z})_0$ = - 0.3 (for Venus).

## Conflict of interest

On behalf of all authors, the corresponding author states that there is no conflict of interest.

Remark regarding contributions of authors as below:

In this research, Dr. Sergey Ershkov is responsible for the general ansatz and the solving procedure, simple algebra manipulations, calculations, results of the article in Sections 1-5 and also is responsible for the search of approximated solutions.

Dr. Alla Rachinskaya is responsible for approximated solving the *non-linear* ordinary differential equation of 2-nd order (15) and (18) by means of advanced numerical methods as well as is responsible for numerical data of calculations and graphical plots of numerical solutions.

Both authors agreed with results and conclusions of each other in Sections 1-7.